\documentclass{amsart}

\usepackage{amssymb}

\usepackage[english,francais]{babel}

\newtheorem{e-proposition}[theorem]{Proposition}

\newtheorem{e-definition}[theorem]{Definition\rm}

\newtheorem{theoreme}{Th\'eor\`eme}[section]

\newtheorem{proposition}[theoreme]{Proposition}
\newtheorem{corollaire}[theoreme]{Corollaire}

\def\og{\leavevmode\raise.3ex\hbox{$\scriptscriptstyle\langle\!\langle$~}}
\def\fg{\leavevmode\raise.3ex\hbox{~$\!\scriptscriptstyle\,\rangle\!\rangle$}}

\def \Z {\mathbb Z}
\def \R {\mathbb R}

\def \N {\mathbb N}

\newcommand{\SpN}{\mathrm{Sp}_{\mathrm{N}}(\R)\,}
\newcommand{\spN}{\mathfrak{sp}_{\mathrm{N}}(\R)\,}
\newcommand{\SN}{\mathrm{S}_{\mathrm{N}}(\R)\,}
\DeclareMathOperator{\supp}{\mathrm{supp}\;}
\DeclareMathOperator{\dd}{\mathrm{d}}
\DeclareMathOperator{\dlO}{d_{\log\, \mathcal{O}}}
\newcommand{\Ho}{H(\omega)}
\newcommand{\Hl}{H_{\ell}(\omega)}
\newcommand{\omO}{\omega^{(0)}}

\title[Anderson \`a potentiel d'interaction g\'en\'erique]{Absence de spectre absolument continu pour un op\'erateur d'Anderson \`a potentiel d'interaction g\'en\'erique}
\author{Hakim Boumaza}
\email{boumaza@math.univ-paris13.fr}
\address{UMR CNRS 7539 - D\'epartement de Math\'ematiques \\
Institut Galil\'ee \\ 
Universit\'e PARIS 13 \\ 
99 avenue Jean-Baptiste Cl\'ement \\ 
93430 VILLETANEUSE \\
FRANCE}

\begin{document}
\maketitle
\begin{abstract}
\selectlanguage{francais}
\noindent Nous pr\'esentons un r\'esultat d'absence de spectre absolument continu dans un intervalle de $\R$ pour un op\'erateur de Schr\"odinger al\'eatoire continu et \`a valeurs matricielles agissant sur $L^2(\R)\otimes \R^N$, pour $N\geq 1$ arbitraire, et dont le potentiel d'interaction est g\'en\'erique dans les matrices sym\'etriques r\'eelles. Pour cela, nous prouvons l'existence d'un intervalle d'\'energies sur lequel a lieu la s\'eparabilit\'e et la stricte positivit\'e des $N$ exposants de Lyapounov positifs de l'op\'erateur. La m\'ethode suivie, bas\'ee sur le formalisme de F\"urstenberg et un r\'esultat de th\'eorie des groupes d\^u \`a Breuillard et Gelander, permet une construction explicite de l'intervalle d'\'energie recherch\'e.
\vskip 5mm

%


\end{abstract}

\selectlanguage{english}
\section*{Abridged English version}

\noindent The question of the Anderson localization remains mostly open for Anderson-Bernoulli models in dimension $d$ higher than $2$. Such an Anderson-Bernoulli model is given  by a family of random operators of the form $H(\omega)=-\Delta + \sum_{n\in \Z^d}\omega_n V(x-n)$, acting on $L^2(\R^d)\otimes \R$, where $V$ is supported in $[0,1]^d$ and the $\omega_n$ are independent and identically distributed (\emph{i.i.d.}) Bernoulli random variables. Since \cite{BK05}, we know that there is exponential localization at the bottom of the almost sure spectrum of $H(\omega)$. In dimension $1$, it is also known (see \cite{DSS02}) that there is localization on every compact interval away from a discrete set of critical energies. In dimension $d\geq 3$, it is commonly conjectured that for high energies, there exist extended states, as for dimension $d=2$ it is conjectured that there is localization at every energies (except maybe a discrete set) like in the one-dimensional case. To tackle the question of the localization in dimension $2$ at all energies, one can start to look at a slightly simpler model of a continuous strip $\R\times [0,1]$ in the plan. We consider the restriction $H_{\mathrm{cs}}(\omega)$ of $\Ho$ to $L^2(\R \times [0,1])$ with Dirichlet boundary conditions on $\R \times \{0 \}$ and $\R \times \{ 1\}$. A possible approach to prove localization for $H_{\mathrm{cs}}(\omega)$ is to operate a discretization in the direction where the strip is bounded, which brings us to consider a quasi one-dimensional operator acting on $L^2(\R)\otimes \R^N$, with $N\geq 1$ an integer.
\vskip 2mm

\noindent In the present article, we look at a particular example of such a quasi one-dimensional operator, denoted by $\Hl$ and defined at (\ref{eq_model}). In the expression of $\Hl$ appears a matrix-valued interaction potential $V$ which is a real symmetric matrix. We prove, at corollary \ref{cor_main}, a result of absence of absolutely continuous spectrum for $\Hl$ in an explicit compact interval $I(N,\ell)$ (for its expression, see (\ref{eq_def_INl})), for a $V$ generic in the sense of the Lebesgue measure. According to Kotani theory, this absence of absolutely continuous spectrum is implied by the non-vanishing of the Lyapunov exponents of $\Hl$, at least out of a set of energies of Lebesgue measure zero. We prove the positivity of the Lyapunov exponents of $\Hl$ at theorem \ref{thm_main}, at least for small enough values of the length parameter $\ell$. Using the formalism of F\"urstenberg, we introduce the tranfer matrices associated to $\Hl$ at (\ref{eq_transfert_mat}) and we shall prove that the F\"urstenberg group $G(E)$ of $\Hl$ (\emph{i.e.} the group generated by the support of the common law of the transfer matrices, defined at (\ref{eq_def_GE})) is equal to the whole symplectic group $\SpN$, for all energies in $I(N,\ell)$, except those in a finite set. According to (\ref{eq_def_GE}), it is sufficient to prove that the group generated by $2^N$ transfer matrices, $<T_{\omO}(E)|\ \omO \in \{ 0,1 \}^N>$, is dense in $\SpN$. Like in \cite{boumazampag} and \cite{boumazampag2}, we use a denseness criterion for finitely generated subgroups of real semisimple connected Lie groups, due to Breuillard and Gelander and stated at theorem \ref{thm_BG}. This criterion gives us the plan of the proof of proposition \ref{prop_GE} which implies theorem \ref{thm_main}. First, we give an explicit form of the transfer matrices, writing them as the exponential of explicit matrices $X_{\omO}(E,V)$ in the Lie algebra $\spN$ (see (\ref{eq_def_X})). Then, we prove that the set $\mathcal{V}$ of interaction potentials $V\in \SN$ such that, for every $E\in \R$, the family $\{ X_{\omO}(E,V)\}_{\omO\in \{0,1\}^N}$ does not generate $\spN$, is of Lebesgue measure zero. But, because a non-zero polynomial has only a finite number of roots, we prove that for $V\notin \mathcal{V}$, there exists a finite set $\mathcal{S}_{\mathrm{V}}$ such that for $E\notin \mathcal{S}_{\mathrm{V}}$, the family $\{ X_{\omO}(E,V)\}_{\omO\in \{0,1\}^N}$ generates $\spN$. Then, for such $V$ and $E$, we construct explicitely the real number $\ell_C$ and the interval $I(N,\ell)$ of theorem \ref{thm_main} such that we have (\ref{eq_prop_INl}). We can finally apply theorem \ref{thm_BG} to the group $\SpN$ to prove proposition \ref{prop_GE}.

\vskip 2mm
\noindent Due to the use of the general criterion on Lie groups, we have been changing the dynamical nature of our problem on Lyapunov exponents to an algebraic problem on generating the Lie algebra $\spN$. This algebraic nature of the objects we are considering allows us to prove a result generic on $V$ and implies the finiteness of the set $\mathcal{S}_{\mathrm{V}}$ (see theorem \ref{thm_main}) of critical energies.

\selectlanguage{francais}
\section{Introduction}\label{secintro}

\noindent La question de la localisation d'Anderson reste largement ouverte pour des mod\`eles d'Anderson-Bernoulli en dimension $d$ sup\'erieure \`a $2$. Un tel mod\`ele est repr\'esent\'e par une famille d'op\'erateurs al\'eatoires de la forme $H(\omega)=-\Delta + \sum_{n\in \Z^d}\omega_n V(x-n)$, agissant sur $L^2(\R^d)\otimes \R$, o\`u $V$ est une fonction \`a support dans $[0,1]^d$ et les $\omega_n$ sont des variables al\'eatoires de Bernoulli ind\'ependantes et identiquement distribu\'ees (\emph{i.i.d.}). Depuis \cite{BK05}, on sait que, pour ces mod\`eles, il y a localisation exponentielle au voisinage du bord inf\'erieur du spectre presque s\^ur de $H(\omega)$. Par contre, on ne sait pas s'il y a ou non localisation dynamique dans cette r\'egion. En dimension $1$, il est connu (voir \cite{DSS02}) qu'il y a localisation dans tout intervalle d'\'energie en dehors d'un ensemble discret d'\'energies critiques. En dimension $d\geq 3$, il est commun\'ement conjectur\'e qu'aux grandes \'energies, il existe des \'etats d\'elocalis\'es, tandis qu'en dimension $d=2$, il est conjectur\'e qu'il y a localisation \`a toutes les \'energies, comme dans le cas unidimensionnel. Pour aborder la question de la localisation en dimension $2$ \`a toutes les \'energies, on peut commencer par s'int\'eresser \`a un mod\`ele l\'eg\`erement plus simple de bande continue $\R \times [0,1]$ dans le plan, en consid\'erant la restriction $H_{\mathrm{bc}}(\omega)$ de $\Ho$ \`a  $L^2(\R \times [0,1])$ avec conditions de Dirichlet aux bords de la bande, $\R \times \{0 \}$ et $\R \times \{ 1\}$. Une approche possible pour tenter de prouver la localisation d'Anderson pour $H_{\mathrm{bc}}(\omega)$ est d'op\'erer une discr\'etisation dans la direction o\`u la bande est de longueur finie, ce qui permet de se ramener \`a l'\'etude d'un op\'erateur quasi-unidimensionnel agissant sur $L^2(\R)\otimes \R^N$, o\`u $N\geq 1$ est un entier.

\vskip 2mm
\noindent A la section \ref{secmodel} de cet article, nous d\'efinirons pr\'ecisement un tel op\'erateur quasi-unidimensionnel, not\'e $\Hl$, et nous pr\'esenterons un r\'esultat d'absence de spectre absolument continu pour celui-ci et ce, pour un potentiel matriciel d'interaction g\'en\'erique. D'apr\`es le th\'eor\`eme R.A.G.E., l'absence de spectre absolument continu dans un intervalle d'\'energies correspond \`a l'absence d'\'etats diffusifs pour ces \'energies. L'absence de spectre absolument continu est une premi\`ere \'etape en vue de prouver la localisation d'Anderson. En vertu de la th\'eorie de Kotani, l'absence de spectre absolument continu pour un op\'erateur de Schr\"odinger quasi-unidimensionnel dans un intervalle $I$ est induite par la non annulation des exposants de Lyapounov de l'op\'erateur pour presque toutes les \'energies dans $I$. Nous allons donc prouver un r\'esultat de s\'eparabilit\'e des exposants de Lyapounov de $\Hl$ dans un certain intervalle $I$. Pour cela, nous allons d\'emontrer que, pour presque tout potentiel matriciel d'interaction, le groupe de F\"urstenberg de $\Hl$ est \'egal \`a tout le groupe symplectique $\SpN$, pour presque toutes les \'energies dans $I$. En cela, nous g\'en\'eralisons \`a un potentiel d'interaction g\'en\'erique, le r\'esultat obtenu dans \cite{boumazampag2}. Comme dans \cite{boumazampag} et \cite{boumazampag2}, nous aurons recours \`a un crit\`ere de densit\'e de sous-groupes dans les groupes de Lie r\'eels connexes semi-simples d\^u \`a Breuillard et Gelander (\cite{BG03}). Ce crit\`ere ram\`ene notre probl\`eme dynamique de stricte positivit\'e des exposants de Lyapounov \`a un probl\`eme alg\'ebrique de g\'en\'eration de l'alg\`ebre de Lie $\spN$. C'est cette nature alg\'ebrique des objets en jeu qui permet de passer du cas particulier \'etudi\'e dans \cite{boumazampag2} au cas g\'en\'erique obtenu ici.

\section{Mod\`ele et r\'esultats}\label{secmodel}

\noindent Nous \'etudions le mod\`ele d'Anderson-Bernoulli suivant :
\begin{equation}\label{eq_model}
H_{\ell}(\omega)=-\frac{\dd^{2}}{\dd x^{2}}\otimes I_{\mathrm{N}}+ V +\sum_{n\in \Z} \left(
\begin{array}{ccc}
\omega_{1}^{(n)} \mathbf{1}_{[0,\ell]}(x-\ell n) & & 0\\ 
 & \ddots &  \\
0 & & \omega_{N}^{(n)} \mathbf{1}_{[0,\ell]}(x-\ell n)\\ 
\end{array}\right)
\end{equation}
agissant sur $L^2(\R)\otimes \R^N$. On suppose que $N\geq 1$ est un entier, que $\ell>0$ est un r\'eel et que $V$ est une matrice sym\'etrique r\'eelle d'ordre $N$ (l'espace de telles matrices est not\'e $\SN$). Pour $i\in \{ 1,\ldots,N\}$, les  $(\omega_{i}^{(n)})_{n\in \Z}$ sont des suites de variables al\'eatoires \emph{i.i.d.}  sur $(\Omega,\mathcal{A},\mathsf{P})$ de loi commune $\nu$ telle que $\{ 0,1\} \subset \supp \nu$ et $\supp \nu$ est born\'e. Pour tout $n\in \Z$, on note $\omega^{(n)}=(\omega_1^{(n)},\ldots,\omega_N^{(n)})$ qui est de loi $\nu^{\otimes N}$. 

\noindent Dans le r\'esultat suivant, la g\'en\'ericit\'e s'entend au sens de la mesure de Lebesgue sur $\SN$ identifi\'ee \`a la mesure de Lebesgue sur $\R^{\frac{N(N+1)}{2}}$.
\vskip 2mm

\begin{theoreme}\label{thm_main}
Pour presque tout $V\in \SN$, il existe un ensemble fini $\mathcal{S}_{\mathrm{V}}\subset \R$ et un r\'eel $\ell_C=\ell_C(N) >0$ tels que, pour tout $\ell \in ]0,\ell_C[$, il existe un intervalle compact $I(N,\ell)\subset \R$ dans lequel les $N$ exposants de Lyapounov positifs $\gamma_1(E),\ldots,\gamma_N(E)$ de $\Hl$ v\'erifient :
\begin{equation}\label{eq_thm_main}
\forall E\in I(N,\ell)\setminus \mathcal{S}_{\mathrm{V}}, \quad \gamma_{1}(E)>\cdots > \gamma_{N}(E)>0.
\end{equation}
\end{theoreme}
\vskip 2mm

\begin{corollaire}\label{cor_main}
Pour presque tout $V\in \SN$, pour tout $\ell \in ]0,\ell_C[$, $\Hl$ n'a pas de spectre absolument continu dans $I(N,\ell)$, presque s\^urement en $\omega$ (o\`u $\ell_C$ et $I(N,\ell)$ sont donn\'es au th\'eor\`eme \ref{thm_main}).
\end{corollaire}

\section{Principe de d\'emonstration des r\'esultats}

\noindent Nous commen\c{c}ons par introduire les matrices de tranfert de l'op\'erateur $\Hl$. Soit $E\in \R$. La matrice de transfert de $\ell n$ \`a $\ell (n+1)$ de $H_{\ell}(\omega)$ est d\'efinie par la relation
\begin{equation}\label{eq_transfert_mat}
\left( \begin{array}{c} 
u(\ell(n+1)) \\
u'(\ell(n+1))
\end{array} \right)= T_{\omega^{(n)}}(E) \left( \begin{array}{c}
u(\ell n) \\
u'(\ell n)
\end{array} \right),
\end{equation}
o\`u $u:\R \to \R^N$ est solution du syst\`eme diff\'erentiel du second ordre $H_{\ell}(\omega)u=Eu$. On introduit alors, pour tout r\'eel $E$, le groupe de F\"urstenberg de $H_{\ell}(\omega)$ :
\begin{equation}\label{eq_def_GE}
G(E)= \overline{<T_{\omO}(E)|\ \omO \in \supp \nu^{\otimes N}>} \supset <T_{\omO}(E)|\ \omO \in \{ 0,1 \}^N>. 
\end{equation}

\vskip 2mm
\begin{proposition}\label{prop_GE}
Pour presque tout $V\in \SN$, il existe un ensemble fini  $\mathcal{S}_{\mathrm{V}}\subset \R$ et un r\'eel $\ell_C >0$ tels que, pour tout $\ell \in ]0,\ell_C[$, il existe un intervalle compact $I(N,\ell)$ tel que :
\begin{equation}\label{eq_GE_SpN}
\forall E\in I(N,\ell)\setminus \mathcal{S}_{\mathrm{V}}, \quad G(E)=\SpN.
\end{equation}
\end{proposition}
\vskip 2mm

\noindent Cette proposition implique le th\'eor\`eme \ref{thm_main} d'apr\`es \cite[Proposition IV.3.4]{BL85}. Pour d\'emontrer la proposition \ref{prop_GE}, nous utilisons le r\'esultat suivant de th\'eorie des groupes d\^u \`a Breuillard et Gelander.
\vskip 2mm

\begin{theoreme}[Breuillard et Gelander, \cite{BG03}]\label{thm_BG}
Si $G$ est un groupe de Lie connexe r\'eel semi-simple, d'alg\`ebre de Lie $\mathfrak{g}$, alors il existe un voisinage de l'identit\'e $\mathcal{O} \subset G$, sur lequel $\log=\exp^{-1}$ est un diff\'eomorphisme et tel que $g_{1},\ldots,g_{m}\in \mathcal{O}$ engendrent un sous-groupe dense dans $G$ si et seulement si $\log(g_{1}),\ldots, \log(g_{m})$ engendrent $\mathfrak{g}$. 
\end{theoreme}
\vskip 2mm

\noindent Ce th\'eor\`eme fait le lien entre une propri\'et\'e topologique de densit\'e d'un sous-groupe engendr\'e par un nombre fini d'\'el\'ements et une propri\'et\'e alg\'ebrique d'engendrer une alg\`ebre de Lie. Il constitue un crit\`ere explicite qui nous donne le plan de la d\'emonstration de la proposition \ref{prop_GE}. Tout d'abord nous calculons explicitement les matrices de tranfert $T_{\omO}(E)$ pour tout $E\in \R$ et tout $\omO\in \{ 0,1\}^N$.

\noindent Posons 
\begin{equation}\label{eq_def_M}
M_{\omO}(E)=V + \mathrm{diag}(\omega_1^{(0)}-E,\ldots, \omega_N^{(0)}-E).
\end{equation}
\noindent Alors, si on note 
\begin{equation}\label{eq_def_X}
X_{\omO}(E,V)=\left( \begin{array}{cc}
0 & I_{\mathrm{N}} \\
M_{\omO}(E) & 0
\end{array} \right),
\end{equation}
on obtient $T_{\omO}(E)=\exp(\ell X_{\omO}(E,V))$.

\noindent L'id\'ee est que, pour $\ell$ suffisamment petit et $E$ pas trop grand, les logarithmes des $T_{\omO}(E)$ valent $\ell X_{\omO}(E,V)$. On regarde donc pour quelles valeurs de $E$ et $V$ la famille $\{X_{\omO}(E,V)\}_{\omO\in \{0,1\}^N }$, not\'ee $\{X_1(E,V),\ldots, X_{2^N}(E,V)\}$, engendre $\spN$.

\noindent Pour $k\in \N^*$, soit 
\begin{equation}\label{eq_def_Vk}
\mathcal{V}_k=\left\{ (X_1,\ldots,X_k)\in (\spN)^k \ |\ (X_1,\ldots,X_k)\ \mathrm{n'engendre}\ \mathrm{pas}\ \spN \right\}.
\end{equation}
Comme engendrer l'alg\`ebre $\spN$ est une condition alg\'ebrique du type non annulation d'une famille finie de d\'eterminants (finie car, pour tout $m\in \N^*$, $\R[T_1,\ldots,T_m]$ est noeth\'erien), il existe $Q_1,\ldots,Q_{r_k}\in \R[(\spN)^k]$ tels que :
\begin{equation}\label{eq_prop_Vk}
\mathcal{V}_k=\left\{ (X_1,\ldots,X_k)\in (\spN)^k \ |\ Q_1(X_1,\ldots,X_k)=0,\ldots, Q_{r_k}(X_1,\ldots,X_k)=0 \right\}.
\end{equation}
Ici, on fait l'identification $\R[(\spN)^k]\simeq \R[T_1,\ldots,T_{k(2N^2 +N)}]$. Soient $E\in \R$ et 
\begin{equation}\label{eq_def_VE}
\mathcal{V}_{(E)}=\left\{ V\in \SN \ |\ \{ X_1(E,V),\ldots,X_{2^N}(E,V)\}\ \mathrm{n'engendre}\ \mathrm{pas}\ \spN \right\}.
\end{equation}

\noindent On montre que $\mathrm{Leb}_{\frac{N(N+1)}{2}}(\mathcal{V}_{(E)})=0$. En effet, soit
\begin{equation}
f_E\ :\ \begin{array}{ccl}
\SN & \to & (\spN)^{2^N} \\
V & \mapsto & (X_1(E,V),\ldots, X_{2^N}(E,V))
\end{array}.
\end{equation} 

\noindent Alors, $f_E$ est polyn\^omiale en les $\frac{N(N+1)}{2}$ coefficients d\'efinissant $V$ et on a :
\begin{equation}\label{eq_prop_VE}
V\in \mathcal{V}_{(E)} \ \Leftrightarrow \ (Q_1\circ f_E)(V)=0,\ldots,(Q_{r_{2^N}}\circ f_E)(V)=0,
\end{equation} 
chaque $Q_i \circ f_E$ \'etant polyn\^omiale en les $\frac{N(N+1)}{2}$ coefficients d\'efinissant $V$. Or, on peut d\'emontrer que, pour tout $E\in \R$, si  $V_0$ d\'esigne la matrice sym\'etrique tridiagonale ayant une diagonale nulle et tous les coefficients de ses sur et sous-diagonales \'egaux \`a $1$, alors $V_0\notin \mathcal{V}_{(E)}$ (voir \cite[Lemme 1]{boumazampag2}). Donc, il existe $i_0\in \{ 1,\ldots, r_{2^N}\}$ tel que $(Q_{i_0}\circ f_E)(V_0)\neq 0$ et, comme la fonction $Q_{i_0}\circ f_E$ est polyn\^omiale et non identiquement nulle,
\begin{equation}\label{eq_prop_Leb_VE}
\mathrm{Leb}_{\frac{N(N+1)}{2}}\left( \{ V\in \SN \ |\ (Q_i \circ f_E)(V)=0) \} \right)=0,
\end{equation}
et, par inclusion,
\begin{equation}\label{eq_prop_Leb_VE_2}
\mathrm{Leb}_{\frac{N(N+1)}{2}}(\mathcal{V}_{(E)})=0.
\end{equation}
Enfin, soit $\mathcal{V}=\cap_{E\in \R} \mathcal{V}_{(E)}$. Alors $\mathcal{V}$ est de mesure de Lebesgue nulle et, si $V\notin \mathcal{V}$, il existe $\mathcal{S}_{\mathrm{V}} \subset \R$ fini tel que, pour tout $E\in \R\setminus \mathcal{S}_{\mathrm{V}}$, $\{ X_1(E,V),\ldots,X_{2^N}(E,V)\}$ engendre $\spN$. En effet, si $V\in \SN \setminus \mathcal{V}$, il existe $E_0\in \R$ tel que la famille $\{ X_1(E_0,V),\ldots,X_{2^N}(E_0,V)\}$ engendre $\spN$. Donc, il existe $i_0\in \{ 1,\ldots, r_{2^N}\}$ tel que $(Q_{i_0} \circ f)(E_0,V)\neq 0$ o\`u :
\begin{equation}\label{eq_def_f}
f\ :\ \begin{array}{ccl}
\R \times \SN & \to & (\spN)^{2^N} \\
(E,V) & \mapsto & (X_1(E,V),\ldots, X_{2^N}(E,V))
\end{array}.
\end{equation}
Or, pour $V$ fix\'e, $E\mapsto (Q_{i_0}\circ f)(E,V)$ est polyn\^omiale et non identiquement nulle, elle n'a donc qu'un ensemble fini $\mathcal{S}_{\mathrm{V}}$ de z\'eros et, pour tout $E\in \R\setminus \mathcal{S}_{\mathrm{V}}$, $(Q_{i_0}\circ f)(E,V)\neq 0$, soit encore, 
\begin{equation}
\forall E\in \R\setminus \mathcal{S}_{\mathrm{V}},\ \{ X_1(E,V),\ldots,X_{2^N}(E,V)\} \notin \mathcal{V}_{2^N}.
\end{equation}

\noindent Nous souhaitons maintenant appliquer le th\'eor\`eme \ref{thm_BG} pour $V\notin \mathcal{V}$ et $E\notin \mathcal{S}_{\mathrm{V}}$. Soient $\lambda_1^{\omega^{(0)}},\ldots, \lambda_N^{\omega^{(0)}}$ les valeurs propres r\'eelles de $M_{\omO}(0)\in \SN$ et soient 
\begin{equation}\label{eq_def_lambda}
\lambda_{\mathrm{min}}=\min_{\omega^{(0)}\in \{ 0,1\}^N} \min_{1\leq i\leq N} \lambda_i^{\omega^{(0)}},\ \lambda_{\mathrm{max}}=\max_{\omega^{(0)}\in \{ 0,1\}^N} \max_{1\leq i\leq N} \lambda_i^{\omega^{(0)}}\ \mathrm{et}\ \delta=\frac{\lambda_{\mathrm{max}}-\lambda_{\mathrm{min}}}{2}.
\end{equation} 
\noindent Si $\mathcal{O}$ d\'esigne le voisinage de l'identit\'e donn\'e par le th\'eor\`eme \ref{thm_BG} pour $G=\SpN$, on pose $\dlO=\max \{ R>0\ |\ B(0,R) \subset \log\, \mathcal{O} \}$,
o\`u $B(0,R)$ d\'esigne la boule de centre $0$ et de rayon $R>0$ pour la topologie induite par la norme matricielle sur $\spN$ induite par la norme euclidienne  sur $\R^{2N}$. Soient
\begin{equation}\label{eq_def_lC}
\ell_C=\min \left( 1,\frac{\dlO}{\delta}\right)
\end{equation}
et, pour $\ell\in ]0,\ell_C[$,
\begin{equation}\label{eq_def_INl}
 I(N,\ell)= \left[ \lambda_{\mathrm{max}}-\frac{\dlO}{\ell}, \lambda_{\mathrm{min}}+\frac{\dlO}{\ell} \right]\neq \emptyset.
\end{equation}
Alors, pour $\ell\in ]0,\ell_C[$,
\begin{equation}\label{eq_prop_INl}
\forall \; \omega^{(0)}\in \{ 0,1\}^N,\ \forall E\in I(N,\ell),\quad \ell X_{\omega^{(0)}}(E)\in \log \mathcal{O}
\end{equation}
et, pour tout $E\in I(N,\ell)$, $\log T_{\omega^{(0)}}(E) = \ell X_{\omega^{(0)}}(E)$, puisque $\exp$ est un diff\'eo\-morphisme de $\log \mathcal{O}$ sur $\mathcal{O}$. Enfin, comme $V\notin \mathcal{V}$, pour $E\in I(N,\ell)\setminus \mathcal{S}_{\mathrm{V}}$, la famille $\{\ell X_{\omO}(E,V)\}_{\omO\in \{0,1\}^N}$ engendre $\spN$. Par le th\'eor\`eme \ref{thm_BG}, pour $E\in I(N,\ell)\setminus \mathcal{S}_{\mathrm{V}}$, $<T_{\omega^{(0)}}(E)\ |\ \omega^{(0)} \in \{ 0,1\}^N>$ est dense dans $\SpN$ et $G(E)=\SpN$ pour ces m\^emes \'energies. Cela d\'emontre la proposition \ref{prop_GE} et par l\`a-m\^eme le th\'eor\`eme \ref{thm_main}. Le corollaire \ref{cor_main} en d\'ecoule en utilisant la th\'eorie de Kotani et Simon sur le spectre absolument continu (voir \cite{KS88}).

\end{document}